
%
\input harvmac
\def\t{\theta}
\def\p{\hfill\break}
\lref\YY{C.N. Yang and C.P. Yang, J. Math. Phys. 10 (1969) 1115.}
\lref\ZamoI{Al.B. Zamolodchikov, Nucl. Phys. B342 (1990) 695.}
\lref\rWO{E. Witten and D. Olive, Phys. Lett. 78B (1978) 97.}
\lref\FS{P. Fendley and H. Saleur, Nucl. Phys. B388 (1992) 609.}
\lref\CFIV{S. Cecotti, P. Fendley, K. Intriligator and
C. Vafa, Nucl. Phys. B386 (1992) 405.}
\lref\FSZ{P. Fendley, H. Saleur and Al.B. Zamolodchikov,
``Massless Flows I: the sine-Gordon and $O(n)$ models'', and ``Massless Flows
II: the exact $S$-matrix approach'', BUHEP-93-8 and 9, USC/93-003 and 4,
LPM-93-07 and 8, hep-th/9304050 and 51.}
\lref\witti{E. Witten, Nucl. Phys. B202 (1982) 253.}
\lref\wittell{E. Witten, Comm. Math. Phys. 109 (1987) 525.}
\lref\pknew{P. Fendley and K. Intriligator, to appear soon.}
\lref\pk{P. Fendley and K. Intriligator, Nucl. Phys. B372 (1992) 533; Nucl.
Phys. B380 (1992) 265.}
\lref\bcov{M. Bershadsky, S. Cecotti, H. Ooguri and C. Vafa,
``Holomorphic Anomalies in Topological Field Theories'', HUTP-93-A008,
hep-th/9302103.}
\lref\lv{A. LeClair and C. Vafa, ``Quantum Affine Symmetry as Generalized
Supersymmetry'', CLNS-92-1150, hep-th/9210009.}
\lref\cvii{S. Cecotti and C. Vafa, ``On Classification of $N$=2 Theories'',
HUTP-92-A064, hep-th/9211097.}
\lref\wz{F. Wilczek and A. Zee, Phys. Rev. Lett. 52 (1984) 2111.}
\lref\mtw{B. McCoy, C. Tracy, and T.T. Wu, J. Math. Phys. 18 (1977) 1058.}
\lref\frac{R. Jackiw and C. Rebbi, Phys. Rev. D13 (1976) 3398;
J. Goldstone and F. Wilczek, Phys. Rev. Lett. 47 (1981) 986.}
\lref\fmvw{P. Fendley, S.D. Mathur, C. Vafa and N.P. Warner, Phys. Lett.
B243 (1990) 257; P. Fendley, W. Lerche, S.D. Mathur and N.P. Warner, Nucl.
Phys. B348 (1991) 66; P. Mathieu and M. Walton, Phys. Lett. 254B (1991) 106.}
\lref\Zamocfn{A.B. Zamolodchikov, JETP Lett. 43 (1986) 730.}
\lref\cv{S. Cecotti and C. Vafa, Nucl. Phys. B367 (1991) 359.}
\lref\ZandZ{A.B. Zamolodchikov and Al.B. Zamolodchikov, Ann.
Phys. 120 (1980) 253.}
\lref\CBS{C. Callias, Commun. Math. Phys. 62 (1978) 213;
R. Bott and R. Seeley, Commun.  Math. Phys. 62 (1978) 235.}
\lref\dvv{R. Dijkgraaf, E. Verlinde, and H. Verlinde, Nucl. Phys.
B352 (1991) 59.}
\Title{\vbox{\baselineskip12pt
\hbox{BUHEP-93-13}}}
{\vbox{\centerline{Exact information in $N$=2 theories }}}
\vglue .5cm
\centerline{Paul Fendley}
\vglue .5cm
\centerline{Department of Physics, Boston University}
\centerline{590 Commonwealth Avenue, Boston, MA 02215, USA}
\centerline{fendley@ryan.bu.edu}
\vglue .5cm
\centerline{\it to appear in the proceedings of SUSY '93}
\vglue 1cm

This is intended to be a simple discussion of work done in collaboration with
S.  Cecotti, K. Intriligator and C. Vafa; and with H. Saleur.  I discuss how
${\rm Tr}\ (-1)^F F e^{-\beta H}$ can be computed exactly in any $N$=2
supersymmetric theory in two dimensions. It gives exact information on the
soliton spectrum of the theory, and corresponds to the partition function of a
single self-avoiding polymer looped once around a cylinder of radius $\beta$.
It is independent of almost all deformations of the theory, and satisfies an
exact differential equation as a function of $\beta$.  For integrable theories
it can also be computed from the exact $S$-matrix. This implies a highly
non-trivial equivalence of a set of coupled integral equations with the
classical sinh-Gordon and the affine Toda equations.
\Date{5/93}

A fascinating example of supersymmetry is $N$=2 supersymmetry in two
dimensions. Initially, its main use was as a world-sheet symmetry in string
theory, but in recent years the subject has taken on a life of its own.  One
reason is that non-renormalization theorems often
make calculations simpler and very elegant.  As a result, one can often do
exact calculations, even in models without conformal symmetry. Here I shall
discuss the ``index''
\eqn\fmf{\tr\ F (-1)^F e^{-\beta H}.}
which is exactly calculable in any two-dimensional $N$=2 theory \CFIV. $F$ is
the fermion number, the conserved charge following from the $U(1)$ symmetry
required in the $N$=2 algebra.

Aside from its intrinsic interest as a calculable non-perturbative quantity,
there are a number of results which follow from the calculation of \fmf:
\p 1)\quad It gives detailed information on the exact soliton spectrum of the
model.
\p 2)\quad It hints at still-unknown mathematical structure
underlying $N$=2 models.
\p 3)\quad It provides a $c$-function on the space of $N$=2 theories.
\p 4)\quad The simplest $N$=2 model describes many properties of
two-dimensional polymers (self-avoiding random walks). The index turns out to
be the partition function for a single such polymer which loops around a
cylinder \FS, yielding the scaling function for the number of such
configurations.
\p 5)\quad In the polymer ``dense'' phase, \fmf\ has poles. These correspond
to places where an excited-state energy level crosses the ground state \FSZ.
Since perturbation theory generally breaks down after such an occurrence, this
gives valuable quantitative information.
\p 6)\quad In any integrable $N$=2 theory, there are two different ways of
doing the calculation, one giving a differential equation, and the other a set
of coupled integral equations. In the simplest case, the differential equation
is the well-known sinh-Gordon equation (a special case of Painlev\'e III); our
result gives a previously-unknown set of integral equations with the same
regular solutions.
\p 7)\quad Reversing the last item, it is the only known example
where the thermodynamic Bethe ansatz equations \refs{\YY,\ZamoI}\ are
equivalent to a differential equation.

To do these $N$=2 calculations, one does not need an infinite-dimensional
symmetry: the calculation of \fmf\ follows only from the finite-dimensional
supersymmetry algebra.  Most exact quantities follow from properties that are
specifically two-dimensional like the infinite-dimensional conformal symmetry.
Since this does not, there is no obvious reason why similar exactly calculable
quantities do not exist in higher dimensions. For example, one can write down
a quantity in four-dimensional $N$=2 self-dual Yang-Mills theory which
generalize many properties of \fmf. Unfortunately, we do not have the tools to
calculate it, but we can hope that this is a technical obstacle and not
something fundamental.

The details of these calculations are in the papers \refs{\CFIV,\FS}. What I
will try to do is describe first the setting: $N$=2 supersymmetric soliton
theories in two dimensions. (These are not a restricted subclass: solitons are
generic in off-critical $N$=2 theories.)  Then I will discuss the two ways of
deriving \fmf, skipping the details but hopefully making the methodology
clear. I will present the results for a specific example, the simplest $N$=2
minimal model perturbed off the critical point.  This is also convenient
because this is the model which allows the calculation of the polymer
quantities. I shall provide a few comments on this equivalence as well.

We first need to look at the $N$=2 symmetry algebra. In two dimensions there
is the fermion-number charge $F$ and four supercharges: the left movers
$Q_{\pm}$ and the right movers $\overline Q_{\pm}$. They obey
\eqn\eSUSY{
\eqalign{&Q^2_+ = Q^2_- = \overline{Q}^2_+ = \overline{Q}^2_- =
\{ Q_+ , \overline{Q}_- \} = \{ Q_- , \overline{Q}_+ \} =0\cr
&\{ Q_+ , \overline{Q}_+ \} = 2\Delta\,\qquad\qquad\{ Q_- , \overline{Q}_- \}
 = 2\Delta^* \cr
&\{ Q_+ , Q_- \} = E -p\qquad\quad\{ \overline{Q}_+ , \overline{Q}_-
\} = E+p ,\cr
&[F,Q_{\pm}]= \pm Q_{\pm}\qquad\qquad\ \ [F,\overline Q_{\pm}]= \mp \overline
Q_{\pm}},}
The crucial ingredient for us is the ``central term'' $\Delta$, which is a
$c$-number, and which depends only on the boundary conditions of the theory.
That is, if the vacua of the theory are indexed by $a$, $\Delta=\Delta_{ab}$,
where $a$ and $b$ are the vacua at left and right spatial infinity,
respectively. One important result of this term is that the mass of a soliton
$m_{ab}$ connecting vacua $a$ and $b$ obeys a Bogomolny bound $m_{ab}\geq
|\Delta_{ab}|$ \rWO. Moreover, the representations of the supersymmetry
algebra are special when this bound is saturated ($m=|\Delta|$).  In this
case, they are two-dimensional; otherwise, they are four-dimensional.  The
``reduced'' representations play a special role in what follows.

First, I need to say what is meant by the word ``index''.  I put it in quotes
because we do not know how to formulate \fmf\ as a index in the precise
mathematical sense. However, like a real one it is independent of many
deformations of the theory. It is well known how Witten's index $\tr\
(-1)^F\exp(-\beta H)$ does not usually depend on $\beta$ or on any finite
deformations of the theory, because excitations with non-zero energy form
fermion-boson pairs whose contributions cancel \witti.  This is actually not
true in general: the arguments of \witti\ only apply in a box with periodic
boundary conditions, which is not always possible with solitons. The fact that
Witten's index in this situation can depend on $\beta$ was discovered some
time ago, and belongs to the class of ``open-space index theorems'' \CBS.

With the open space required for solitons, we can show that the index \fmf\
depends only on ``$F$-term'' perturbations of the action (including $\beta$);
it does not depend on ``$D$-terms''. This $F$-term dependence is not a
nuisance, it is the point: the differential equations for the index are in
terms of the parameters which describe the $F$-terms.  $D$-terms have all four
supercharges operating on some field (in superspace this is an integration
over $d^4\theta$), while $F$-terms have only $Q_+$ and $\overline Q_+$
($d^2\theta^+$) or the complex conjugate $Q_-$ and $\overline Q_-$
($d^2\theta^-$). The powerful theorems I mentioned earlier say that quantum
effects renormalize $F$-terms only by their naive scaling dimension. In a
Landau-Ginzburg theory, the $D$- and $F$-terms correspond to the kinetic term
and the superpotential, respectively, so the theorems say that the form of the
superpotential is not renormalized.  The index is independent of $D$-term, so
it does not get renormalized. It changes only under perturbation by relevant
and marginal operators, and there are generically only a finite number of
these.  In a sigma model, deformations of the target-space metric which do not
change the K\"ahler class do not change the $F$-term; in the case with target
space a sphere ($CP^1$) this means that the index depends only on the area of
the sphere.

To prove that our index does not depend on the $D$-terms requires only the
fact that $\{(-1)^F,Q\}=0$ for any $Q$, the relations \eSUSY, and the cyclic
property of the trace. Any $D$-term variation of the action can be written as
$\{Q_+,[\overline Q_-,\Lambda]\}$, where $\Lambda$ is some field (in fact
$\Lambda$ is $Q_- \overline Q_+$ of something). Thus infinitesimally the
change in \fmf\ is
\eqn\Dproof{\eqalign{ \delta I=&\tr  F(-)^F
 \{Q_+,[ \overline Q_-,\Lambda]\} e^{-\beta H}\cr
=&\tr  F(-)^F \left( Q_+ [ \overline Q_-,\Lambda]
 +[ \overline Q_-,\Lambda] Q_+\right) e^{-\beta H}\cr
=&\tr  (-)^F
[F, Q_+] [ \overline Q_-,\Lambda] e^{-\beta H}\cr
=&\tr  (-)^F
\left(Q_+\overline Q_- \Lambda- Q_+\Lambda\overline Q_-\right) e^{-\beta H}\cr
=&\tr  (-)^F  \{ Q_+ ,\overline Q_-\}\Lambda e^{-\beta H}\cr
=&0\ .\cr}}
If you try a similar argument with an $F$-term variation, you find that the
index does indeed change. Similarly, objects like $F^2 (-1)^F$ do depend on
the $D$-term.  This argument is heuristic, because one must define quantum
operators precisely to ensure the cyclic property of the trace. One can also
prove the $D$-term invariance by using the path integral; I leave it as an
exercise to the reader to decide if this is more or less rigorous than the
above argument.

Before computing the index exactly, it is first useful to calculate it in some
simple limits. In the high-temperature (ultraviolet) limit and in fact at any
critical point, its largest eigenvalue turns out to be proportional to the
central charge of the conformal theory.  (Notice that the index is actually a
matrix, with rows and columns labelled by the different vacua, corresponding
to the allowed boundary conditions at plus and minus spatial infinity.) Thus
the index provides a $c$-function on the space of $N$=2 theories, although it
is not the same as Zamolodchikov's \Zamocfn. Although in all known unitary
examples the index decreases as one goes into the IR, we do not have a general
proof of this.

For low temperature (the infrared), we can expand it in powers of $\exp (-m
\beta)$, where $m$ is any particle mass. This corresponds to an expansion in
terms of particle states.  It is easy to calculate the one-particle piece
explicitly. We need only the density of states for a single particle, which
follows from putting the particle in a box of length $L$ and quantizing its
momentum via $p=n\pi /L$. The density of states is $g(p)\equiv dn/dp =
L/\pi$. The contribution of a particle with fermion number $f$ and mass $m$ to
the index is
\eqn\onep{\eqalign{&f (-)^f\int_0^\infty dp\ {L\over\pi}\
e^{-\beta\sqrt{p^2 +m ^2}}\cr
=&f (-)^f{L\over\pi} K_1 (m\beta),\cr}}
where $K_1(x)$ is a Bessel function.  We must not forget that the particles
come in multiplets under the supersymmetry. For particles in the reduced
doublet representation, the multiplet has charges ($f$,$f$--1) and so the
contribution is proportional to $f-(f-1) =1$ and is
$$ (-)^f{L\over\pi} K_1 (\Delta\beta).$$
I have written the mass $m$ of this reduced multiplet as $\Delta$ to show that
this contribution is purely global: it depends only on the boundary
conditions. I would also like to point out that $f$ is usually not an
integer in these $N$=2 theories \pk; the phenomenon of fractional fermion
number in soliton theories has been known for some time \frac.  For a
four-dimensional multiplet ($f$+1,$f$,$f$,$f$--1), something interesting
happens: the one-particle contribution vanishes! This is because $f$+1 --$2f$ +
$f$--1 =0. We see the first useful piece of information coming from the
index: it counts the number of solitons in reduced multiplets. Since one
method of exact computation comes only from the $N$=2 chiral ring, this is a
valuable way of finding the (doublet) spectrum of the theory.  We also see how
the index is a much simpler object than the full partition function, but still
complex enough to give us non-trivial information.

One can calculate the two-particle contribution in this manner, but beyond
this (except for the integrable theories to be discussed shortly) the
calculation is too hard. One might think that all four-dimensional
representations of supersymmetry do not contribute to the index, by the above
argument. However, in the multi-particle case, the argument is not applicable.
If space is a finite box, one cannot take periodic boundary conditions and
keep the solitons. Any other boundary condition breaks supersymmetry, so the
members of a multiplet do not necessarily cancel among one another.  If space
is the infinite line, one has a continuum of states. The densities for the
members of a multiplet are not necessarily the same, so again no cancellation
need happen. This is why Witten's index can depend on $\beta$ in soliton
theories. One can think of this as sort of an anomaly: by putting the theory
in a box one breaks the supersymmetry, and the effect remains even as one
takes the box boundaries to infinity.

We now discuss the methods for computing the index exactly. The first method
uses the exact $S$-matrix of the solitons. This can be found in an integrable
theory, which, luckily, covers many of the simplest and most interesting $N$=2
theories \refs{\fmvw,\pk}. Understanding how to get an exact $S$-matrix is a
long story, but the constraints of an integrable theory generally fix it
uniquely: for a discussion see \ZandZ. {}From the exact $S$-matrix, one can
then use the thermodynamic Bethe ansatz to calculate the index
\refs{\YY,\ZamoI}. This is a clever trick: one puts the particles in a box and
quantizes the momenta like we did above for the one-particle contribution.
Because the exact $S$-matrix is completely elastic (individual momenta do not
change in a collision) the $i$th momentum is quantized via the condition
\foot{This relation is for periodic boundary conditions, but is easily
modified to the fixed case required for solitons.}
$$e^{i p_i L} \prod_j S(p_i,p_j) =1\ ;$$
one can think of this as bring the $i$th particle around the world and
scattering it off of all the others. If the scattering is trivial ($S$=1) we
obtain the free-particle quantization condition $p_i=2\pi n_i /L$.  Using this
quantization as a constraint, one minimizes the free energy to obtain
thermodynamic quantities like \fmf. The answer is given in terms of non-linear
integral equations. The calculation was done for many $N$=2 theories in \pk; I
will give an example below.

The other method applies to any $N$=2 theory, and uses
``topological-antitopological fusion'' \cv. The input required here is the
chiral ring coefficients $(C_i)^j_k$, which can be determined using
topological field theory \dvv. First, we define a ``topological'' state
$|a\rangle$ on a disk by adding a term to the action $$\int {i\over 2} j_{\mu}
\omega^{\mu}.$$ This couples the fermion-number current $j_{\mu}$ to the spin
connection of the manifold $\omega^{\mu}$. Basically, this amounts to putting
a (Ramond) ground state $a$ on the boundary of the disk. If we wanted to do
topological field theory on the sphere, we would then glue another disk like
this to it, and calculate the topological metric $\eta_{ab}$. This has all
sorts of marvelous properties: for example, it is independent of the shape of
the sphere.  However, we are interested in information in the full theory, not
the topological subsector. By putting the ground state $|a\rangle$ at one end
of a cylinder of circumference $\beta$ and long length $L$, and a conjugate
state $\langle \overline b |$ at the other end (this has fermion-number
current coupled to {\it minus} half the spin connection), we calculate the
ground-state metric $g_{a\overline b}$.  This can be thought of as a
generalization of Berry's phase to a curvature on the multi-dimensional space
of ground states; this was first discussed in quantum mechanics in \wz.  In
\cv, a differential equation for $g_{a\overline b}$ was derived. The
derivatives are in parameter space, which here governs the $F$-term variations
(the relevant and marginal $N$=2-preserving perturbations of the theory, which
are in one-to-one correspondence with the ground states). Letting $i$ and $j$
index these variations, the (matrix) equation is
\eqn\forg{\bar \partial_j (g \partial_i g^{-1})=
\beta^2[C_i, g C_j^\dagger g^{-1}]}
It is a simple calculation to derive that the index \fmf\ is
\eqn\ttind{I= iL (g \partial_{\beta} g^{-1} + {n\over\beta}),}
where $n$ is the coefficient of the chiral anomaly (in a Landau-Ginzburg
theory it is the number of superfields) and $\partial_{\beta}$ means that we
vary the inverse temperature $\beta$ or equivalently the mass scale of the
theory (since all quantities must depend on the dimensionless ratio $m\beta$).
This should not be surprising: we have coupled fermion number to the geometry,
so when one varies $\beta$ and changes the circumference of the cylinder, it
brings down the $F$ in front of $\tr F(-)^F \exp(-\beta H)$.

We will display these results in the simplest $N$=2 model: the first member of
the $N$=2 discrete series perturbed by the only relevant operator which
preserves supersymmetry. This turns out to be the ordinary sine-Gordon model
at a specific coupling ($\beta^2={2\over 3} 8\pi$) where it is $N$=2
supersymmetric. In the Landau-Ginzburg picture this is described by a single
superfield $X$ (consisting of a complex boson $\phi$ and a Dirac fermion) with
superpotential $W=X^3/3- \lambda X$. The potential for $\phi$ is $|\partial
W/\partial X|^2 {}_{X=\phi}$, giving $(\phi^2 -\lambda)(\phi^{*2}-\lambda^*)$
(supersymmetric $\phi^4$ with a double well). This has two minima, at
$\phi=\pm \sqrt{\lambda}$. The entire spectrum of this theory turns out to be
two doublets of solitons. One doublet has $\phi =-\sqrt{\lambda}$ at negative
spatial infinity and $\phi =\sqrt{\lambda}$ at positive infinity, while the
other goes in the opposite direction. Since the theory has a charge
conjugation invariance the fermion numbers of a doublet must be $(1/2,-1/2)$.
The model is integrable and the exact $S$-matrix for these solitons was found
in \pk.

We can now apply our two methods. The result of the first method (starting
from the exact $S$-matrix) is that the index is
\eqn\piiiF{Q(z=m\beta)\equiv -i{\beta\over L} I= z\int {d\t \over 2\pi}\cosh
\t e^{-A(\t ;z)}.}
where the function $A(\t;z)$ obeys the integral equations
\eqn\piiiint{\eqalign{
A(\t ;z)&=z\cosh \t -\int {d\t '\over 2\pi }{1\over \cosh (\t -
\t ')}\ln (1+B^2(\t';z))\cr
B(\t ;z)&=-\int{d\t '\over 2\pi }{1\over \cosh (\t -\t ')}e^{-A(\t ';z)}.}}

The topological-antitopological analysis of this theory was discussed at
length in \cv .  The metric on the space of ground states in the basis spanned
by $1$ and $X$ can be written as $g=\beta e^{\sigma _3u(z)/2}$. Using
\forg\ with the chiral ring $X^2=\lambda$, it follows that $u(z)$ satisfies
the radial sinh-Gordon equation, (a special case of Painlev\'e III)
\eqn\piii{   {d^2u \over dz^2}+{1\over
z}{du\over dz}=\sinh u.}
The index $Q(z)$ is
\eqn\piiiQ{Q(z)={1\over 2}z{d\over dz}u(z).}
The boundary conditions are fixed by demanding regular behavior at
$mR\rightarrow\infty$, and have been discussed in detail in \mtw.

The equivalence between solutions of these integral equations and this
differential equation was unknown and is very surprising, since both types of
equation have been studied extensively. The sinh-Gordon equation of course is
well known in mathematics and physics, while as a result of the thermodynamic
Bethe ansatz, integral equations like \piiiF\ have been studied a great deal
in the last several years. We lack a direct mathematical proof of the
equivalence, although we have checked it numerically and in high- and
low-temperature limits. We have a hint that a great deal of structure in these
models is still undiscovered. I would also like to comment that for the
remainder of the $N$=2 discrete series perturbed by the least-relevant
operator and for the $CP^1$ sigma model, the index obeys the same differential
equation, but with a slightly different boundary condition.  The integral
equations are modified slightly by adding a constant to the first of \piiiint.
If one perturbs the minimal model by the most-relevant operator, one obtains
the Toda hierarchy of differential equations. One then can find integral
equations equivalent to these as well.

The last thing I would like to discuss is the connection with polymers \FS.
It turns out that many properties of 2d polymers can be described by studying
this $N$=2 theory.  Unfortunately, we lack a direct map of all quantities in
the polymer theory to those in the $N$=2 theory; the equivalence is often
quite subtle. However, we have shown that the index corresponds to the
partition function of a single ring polymer (closed self-avoiding random walk)
looped once around a cylinder of radius $\beta$. The parameter $m$ is a scaled
version of the distance from the polymer critical point. This critical point
separates the high-temperature ``dilute'' phase from the low-temperature
``dense'' phase. In the dense phase, the polymers cover all of space.  This
corresponds to $(m\beta)^{4/3}$ negative, and the theory is non-unitary.  The
integral equations derived above diverge when continued to these values, but
the differential equation can be continued, and becomes the cosh-Gordon
equation. When one continues the solution appropriate in the dilute phase, one
finds poles in the dense phase. In \FSZ, we have argued in a number of ways
that these poles correspond to level crossings.  These are places where an
excited-state energy reaches zero and crosses the high-temperature
(supersymmetric) ground state. There are an infinite number of these crossings
between the polymer critical point and the zero-temperature dense critical
point. One can attempt to find an exact $S$-matrix for the dense phase as
well, but because of the level crossings, the issue is very subtle.

There are a number of related developments:
\p 1) Some new models of massless solitons \pknew\
describe the flows from $c$=3 into an $N$=2 minimal model with central charge
$3k/k+2$. {}From the exact $S$-matrix, we calculate the index, and find that it
does not change during this flow.  Thus the flow is all $D$-term! This can
be thought of as a flow in the Landau-Ginzburg model from a free superfield at
$c$=3 to the (unknown) $D$-term which describes the minimal model. All along
the way, the potential stays at $X^{k+2}$.
\p 2) There is a similar but distinct object, the elliptic genus  \wittell
$$\tr\ e^{i\alpha F_L} (-1)^{F_R} q^{L_0} \bar q^{\bar L_0},$$
which also does not depend on the $D$-term. As compared to our
index, this has the advantage that one computes it on a torus (not just the
cylinder) and gets a full character. The disadvantage is that it requires
separate left and right fermion numbers, which usually does not happen off the
critical point. Thus it is a function of $\beta/L$, while \fmf\ is a function
of $m\beta$.
\p 3) Since $N$=2 supersymmetry is a simple case of the affine quantum-group
symmetry underlying many two-dimensional models, one can hope that similar
objects occur in non-supersymmetric two-dimensional models. Some preliminary
steps in this direction were made in \lv.
\p 4) This work has been extended to show how to relate the soliton spectrum
off the critical point to the charges of the chiral primary fields at the
critical point \cvii. This observation may be useful in classifying $N$=2
conformal field theories. In addition, similar objects have been discussed in
\bcov; these have deep geometrical and physical significance, and show that the
uses of $N$=2 supersymmetry are far from being completely known!

\bigskip\bigskip
\centerline{\bf Acknowledgements}
\bigskip

I would like to express my deep gratitude to my collaborators in the $N$=2
racket: K. Intriligator, S. Cecotti, W. L\"erche, S. Mathur, H. Saleur, C.
Vafa, N. Warner, and Al.  Zamolodchikov, for their efforts and for their
insight.  This work was supported by DOE grant DEAC02-89ER-40509.

\listrefs
\end